\documentclass{aa}
\usepackage{graphicx}

\def\fmag{\hbox{$.\!\!^m$}}  
 
\newcommand{\mincir}{\raise
-2.truept\hbox{\rlap{\hbox{$\sim$}}\raise5.truept\hbox{$<$}\ }}
\newcommand{\magcir}{\raise
-2.truept\hbox{\rlap{\hbox{$\sim$}}\raise5.truept\hbox{$>$}\ }}
\newcommand{\minmag}{\raise
-2.truept\hbox{\rlap{\hbox{$<$}}\raise6.truept\hbox{$<$}\ }}

\begin{document} 

 
\title{Physical Properties of Hickson Compact Groups and of the Loose Groups
  within which they are Embedded}

\author{Hrant Tovmassian \inst{1,2}
\and
Manolis Plionis\inst{3,1}
\and
J.P. Torres-Papaqui\inst{1}
}

\offprints{H.M.Tovmassian} 
 
\institute{Instituto Nacional de Astrof\'isica Optica y Electr\'onica, AP
51 y 216, C.P. 72000, Puebla, Pue., Mexico, e-mail: hrant@inaoep.mx,papaqui@inaoep.mx \\
\and
Sternberg Astronomical Institute, Moscow State University, Moscow, Russia\\
\and
Institute of Astronomy \& Astrophysics, National Observatory of Athens, I.Metaxa 
\& B.Pavlou, P.Penteli 152 36, Athens, Greece, e-mail:mplionis@inaoep.mx
}

\date{Received ... 2005/ Accepted .... 2005} 

\authorrunning{H. M. Tovmassian, M. Plionis, J. P. Torres-Papaqui}
\titlerunning{Hickson Compact Groups and their Environment}
\maketitle 

\begin{abstract}
Using new data and an enlarged group sample we verify some of our
previously
published results and present a number of new facts that suggest that the 
compact groups could be casual concentrations in prolate-like looser
groups and thus the
nature of compact and ordinary poor groups should probably be the same.

To this end we use the Sloan Digital Sky Survey (SDSS) redshift catalogue to
look for galaxies with accordant redshifts in the nearby environment 
(up to $\sim$2 Mpc) of 15 Hickson Compact Groups (HCG). 
We also use known member redshifts of
looser groups in the environment of 7 other HCGs. 
Using this sample of 22 HCGs we find that: (a) HCG's tend to be
aligned with the overall galaxy distribution in their $\sim$1 Mpc
environment, (b) the well-established 
orientation effect by which the group velocity dispersion, $\sigma_v$, 
correlates with group 
axial ratio $q$, is present and particularly strong also in the HCG+environment
 systems, (c) the radial velocity dispersion $\sigma_v$ of the HCG+environment
systems as well as of ordinary poor groups only weakly depends on the 
group richness, i.e. on the mass, (d) the mean absolute magnitude 
$\langle M_K \rangle$ of E/S0 galaxies in HCGs is similar to the 
corresponding one in ordinary poor groups, and is brighter than that 
of isolated E/S0's, indicating that they were formed by the merging of 
two galaxies of similar luminosity, (e) the fraction of E/S0 galaxies 
in these HCGs depends, albeit weakly, on the group richness and 
on $\sigma_v$, (f) the fraction of AGNs is similar in
the HCGs and their close environment, while the
fraction of starburst galaxies is significantly higher in the HCGs,
(g) the fraction of active galaxies (AGNs
and starbursts) is anti-correlated with the velocity dispersion of the
HCG+environment systems.

The combination of all the above facts constitutes a picture in which
compact groups are condensations within looser prolate-like elongated
systems and they appear as compact when their member galaxies, moving in 
radial orbits along the group elongation, happen to come close to each
 other (in which case dynamical interactions among these galaxies
become even more probable) 
or when the group is oriented close to the line of sight, so that many
of its members are projected over a small solid angle. 
The probability of either case is small and therefore the number of 
CGs should be much smaller than that of ordinary groups, as it is observed.

Furthermore, the observed fractions of early-type and active galaxies
as well as their correlations with the group velocity dispersion suggests
a picture by which nuclear activity and galaxy transformation by
merging is instigated by effective gravitational interactions in the 
low-velocity dispersion groups, which then dynamically evolve 
via virialization processes to higher velocity dispersion groups, 
which thus have a higher fraction of early-type galaxies.

\end{abstract}

\section{Introduction}
Compact Groups are environments with high surface galaxy density and
relatively low radial velocity dispersion, $\sigma_v$. 
In recent decades several lists of CGs have been compiled (see the
review by Hickson 1997; Focardi \& Kelm 2002; Iovino et al. 2003; Lee
et al. 2004; de Carvalho et al. 2005) but still the most extensively 
studied sample of 
CGs, is that of Hickson Compact Groups (HCGs) (Hickson 1982). 
N-body simulations have showed that strong galaxy interactions and merging 
should occur frequently in such systems, and CGs should eventually
evolve to form a single elliptical galaxy 
on a time scale of a few orbital periods (Barnes 1985, 1989, 1990; Mamon 1987; 
Bode et al. 1993). For this reason the very existence of CGs 
has been questioned and debated. To explain the existence of CGs it has been 
suggested that field galaxies fall from time to time onto CGs, thus 
keeping the number of groups approximately constant 
(Diaferio et al. 1994, 1995; Governato et al. 1996; 
Ribeiro et al. 1998). However, observational estimates (Zepf et
al. 1991; Moles et al. 1994, Zepf 1993; Mendes de Oliveira 
\& Hickson 1994; Coziol, Brinks \& Bravo-Alfaro 2004) 
suggest relatively low merger rates. 
To explain these observations Coziol, Brinks \& Bravo-Alfaro
(2004) put forward two hypotheses: either the evolution of 
galaxies accelerates in richer systems, or CG's form earlier in
massive structures than in lower mass systems.
Solutions to the problem come also (a) from the fact that CGs are 
small subsystems in larger ordinary (hereafter, loose) groups (LGs) 
(Rose 1977, Sulentic 1987; Rood \& Williams 1989; West 1989; Mamon 1990;
Vennik et al. 1993; Diaferio et al. 1994; Rood and Struble 1994; Tovmassian \& 
Chavushyan 2000; Tovmassian 2001; Tovmassian, Yam, \& Tiersch 2001)
and (b) from a sufficiently high formation rate, so that despite their
short merging times they are still abundant enough to be observed at
the present numbers (e.g. Mamon 2000). Ramella et al. (1994) concluded 
that we see HCGs because they are being continually formed in
collapsing LGs, as it was predicted by some N-body simulations 
of rich groups of galaxies (eg. Diaferio et al. 1995). 

It has been shown that HCGs, Shakhbazian Compact Groups 
(Shakhbazian 1973; Baier \& Tiersch 1979 and references therein) 
as well as poor groups have a prolate-like shape (Malykh \& Orlov 1986; 
Oleak et al. 1995; Orlov et al. 2001; Plionis, Basilakos \& Tovmassian 2004). 
Tovmassian, Martinez, \& Tiersch (1999), showed that for HCGs the 
velocity dispersion correlates with the group projected shape: 
the lower the group projected axial ratio the smaller its
velocity dispersion. They concluded that member galaxies move in
radial orbits predominantly along the group elongation. Furthermore,
Tovmassian (2001), Tovmassian, Yam, \& Tiersch (2001), Tovmassian, \& 
Tiersch (2001) showed that members of LGs in the environment of CGs
 are distributed in narrow strips of a couple of hundred kpc width, 
and of about 1 Mpc length while the projected position angle of the LG 
generally coincides with that of the CG. Using galaxy redshifts 
in the environments of 6 HCGs 
of de Carvalho et al. (1994) and Zabludoff \& 
Mulchaey (1998), Tovmassian \& Chavushyan (2000) showed that 
accordant redshift galaxies in the environments of these 
HCGs and poor groups obey even more strongly the dependence 
of the projected shape on the $\sigma_v$ of the whole system. It follows 
that members of LGs should be gravitationally bound with CGs and have 
quasi-regular movement predominantly along the elongation of the
 whole system.  Tovmassian \& Chavushyan (2000), Tovmassian (2002)
 put forward the idea that CGs are ordinary poor groups in which 
the brighter members happen to be close to each other by chance
during the orbital movement around the centre of gravity, 
or if the groups are seen end-on and most of its members are projected 
near to each other on the sky.

In this paper we present indications for a possible common 
nature of compact and loose groups of galaxies.
We show that CGs are condensations in LGs and that the HCG+environment
systems obey similar correlations as loose groups.  

\section{The sample}
In order to study the dynamics of HCG+environment systems we analyse a 
sample of 22 HCGs. 
Since the projected mean length of poor groups is 
generally less than 1 Mpc\footnote{Throughout this paper we use 
$H{_0}=72$ km s$^{-1}$ Mpc$^{-1}$} (e.g. see Plionis, Basilakos, 
\& Tovmassian 2004), we look for accordant redshift
galaxies up to a distance of $\sim$ 2 Mpc from the original HCGs.
We use mainly the SDSS (Abazajian et al. 2005) to search 
for galaxies with redshift separation from the centre of the
corresponding HCG of $\delta (cz) = 1000$ km s$^{-1}$ and within 
a circle of projected radius $\sim$2 Mpc radius. In the area covered by the SDSS we
find 22 HCGs out of which 15 ($\#$ 7, 25, 35, 43, 45, 49, 50, 56, 58, 
60, 66, 68, 82, 88 and 98) are sampled out to the projected separation
limit discussed previously, while the other 7 are
located near the SDSS survey limits and thus are not sampled adequately.
Accordant redshift galaxies were found in the 
environment of 14 of these 15 groups (see Fig. 1). We have noted that in
most cases the magnitude distribution of the accordant redshift
galaxies does not differ from that of the HCG galaxies.

Furthermore, the environment of 6 other HCGs (HCG~40,
HCG~63, HCG~64, HCG~67, HCG~87 and HCG~97) have been studied 
spectroscopically up to sufficiently large projected distances from
their centre (de Carvalho et al. 1997; Zabludoff \& Mulchaey 1998). 
The environment of one more group,
HCG~92 (the well known Stephan Quintet), has been studied in HI 
emission by Shostak, Sullivan III, \& Allen (1984) who detected 5 accordant redshift 
galaxies, and by Williams, Yun \& Verdes-Montenegro (2002)
who detected one more accordant redshift galaxy (last image in Fig 1). We assume 
that the discordant redshift galaxy NGC 7320 is not a member of the group.
\begin{figure}
\includegraphics[scale=0.8]{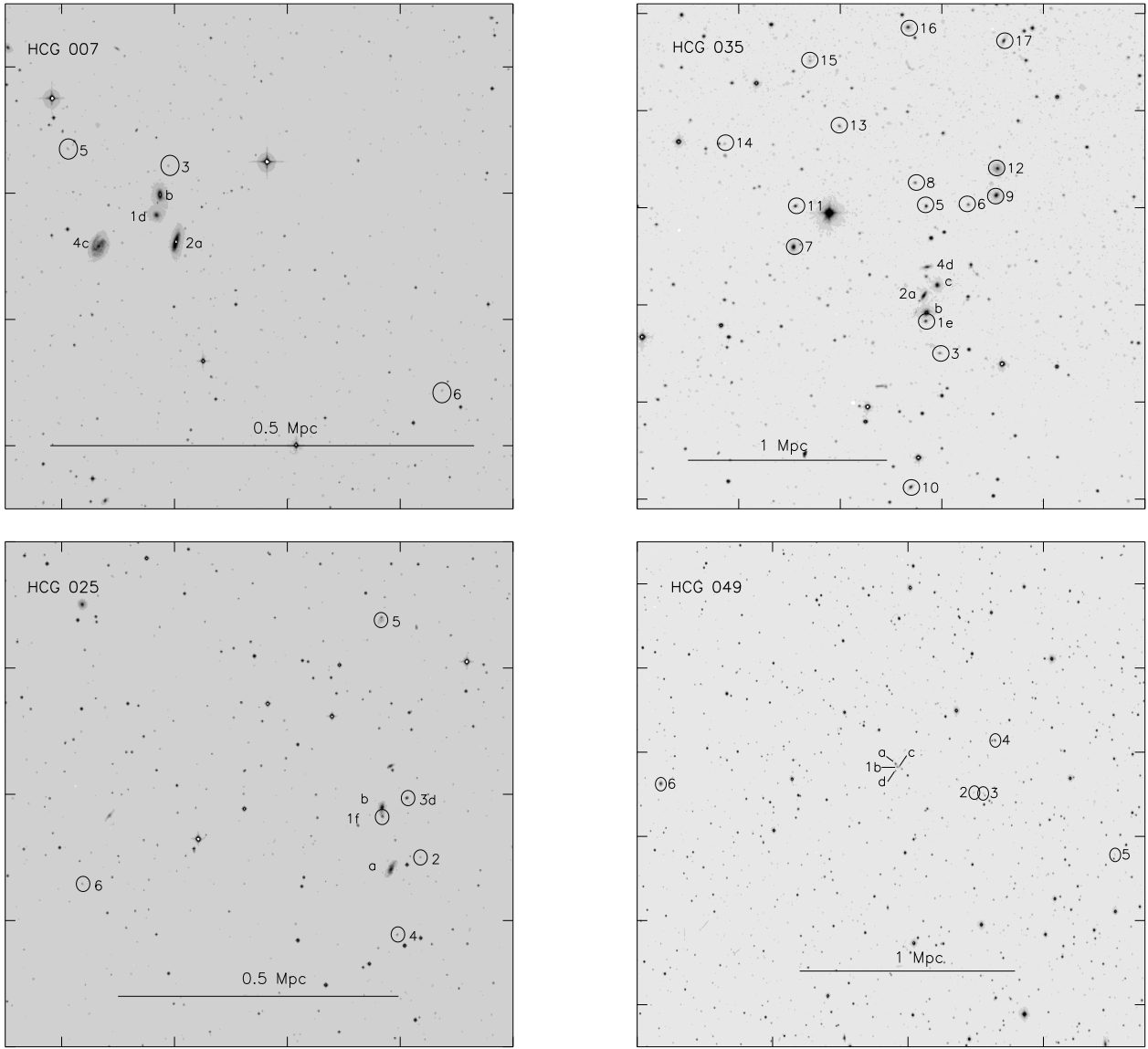}
\caption{Digital Sky Survey images of the environment of some of our HCGs. 
Original HCG members are marked by letters ($a$, 
$b$ etc.), as in Hickson (1994) while the SDSS galaxies are marked by
circles and they are numbered (1, 2, 3, etc). Accordant redshift
galaxies are those marked only by a number.}
\label{fig1}
\end{figure}

\begin{figure}
\includegraphics[scale=0.8]{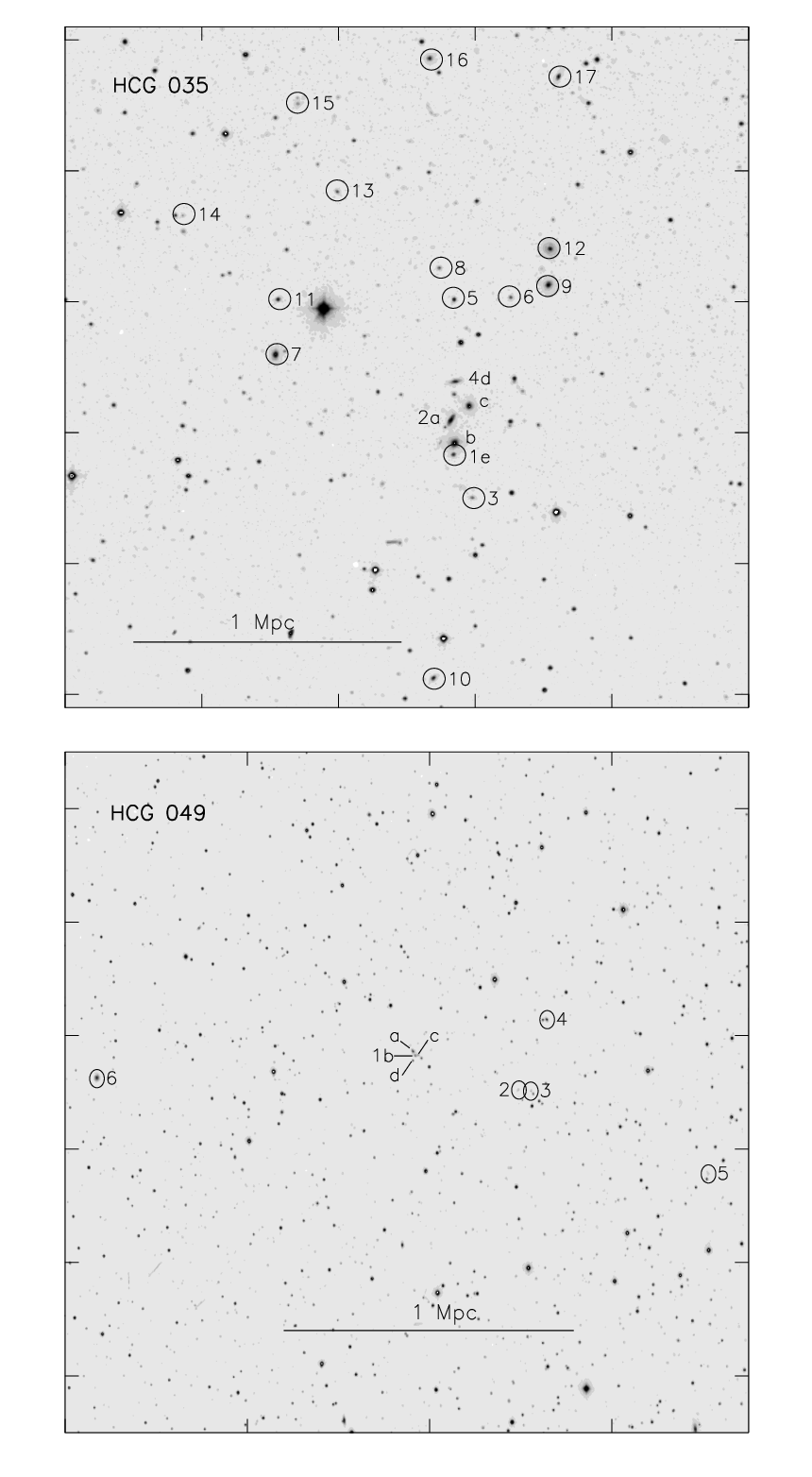}\\
Fig. 1. -- Continuation
\end{figure}

\begin{figure}
\includegraphics[scale=0.8]{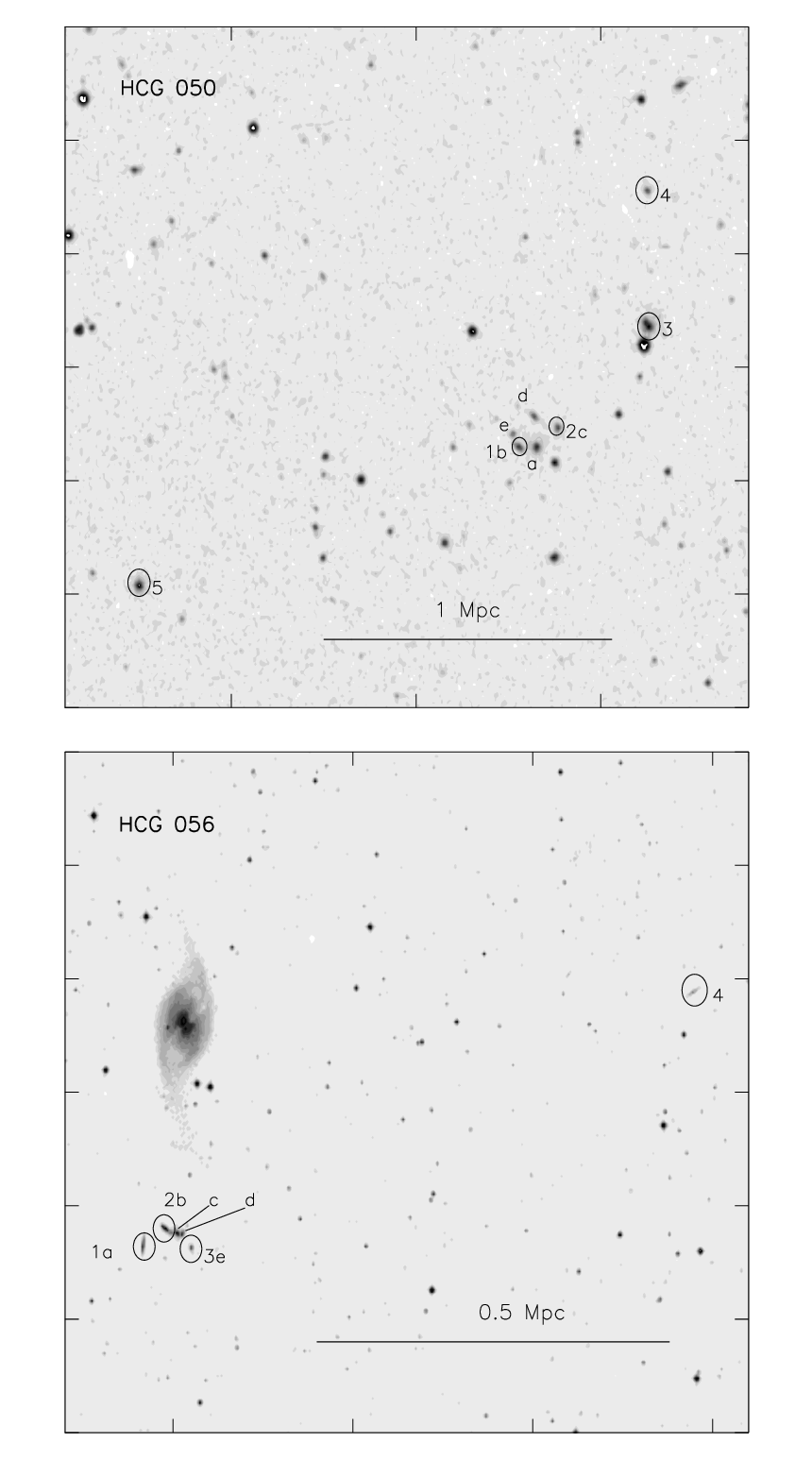}\\
Fig. 1. -- Continuation
\end{figure}


\begin{figure}
\includegraphics[scale=0.8]{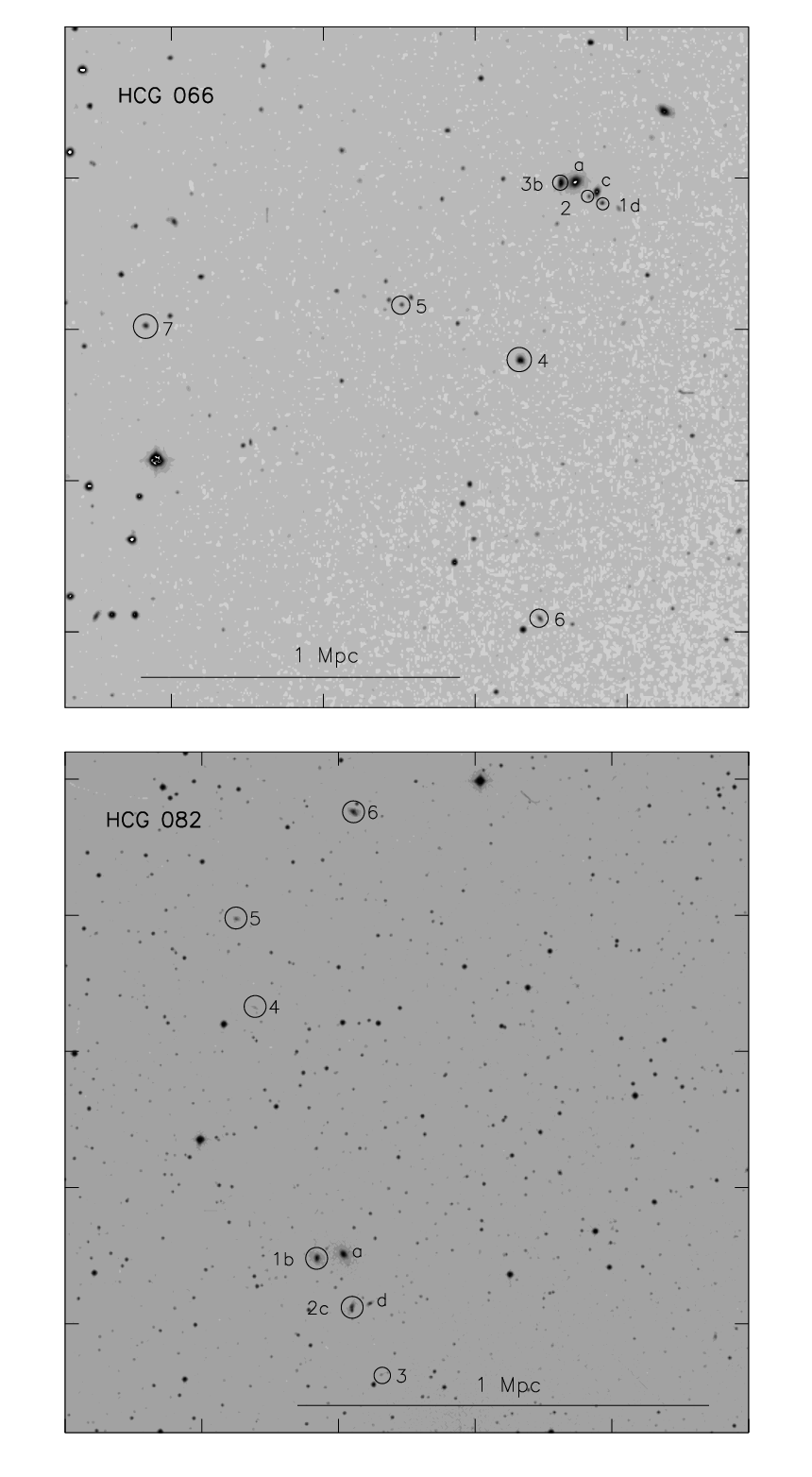}\\
Fig. 1. -- Continuation
\end{figure}



We found that the velocity dispersions of a few HCG+environment
 systems are higher than those of loose groups which generally do not 
exceed 400 km s$^{-1}$.
Unusually high are the $\sigma_v$'s of 
groups HCG~50 and HCG~82: 546 km s$^{-1}$ and 468 km 
s$^{-1}$, respectively. Note that using the original four members of
HCG~82 we obtain an even larger value: $\sigma_v \simeq 620$ km s$^{-1}$. 
This leads us to the suspicion that they could be the result of
projection effects.
The group HCG 50 could well be the superposition of two independent 
groups. The original four galaxies, $a$, $b$, $c$ and $e$ (Hickson 
designations) of this group could compose two groups. 
One with $\langle cz \rangle=41520$ km s$^{-1}$ and 
$\sigma_v$=263 km s$^{-1}$ (HCG 50a), and one with $\langle cz \rangle
=42450$ km s$^{-1}$ and $\sigma_v=96$ km s$^{-1}$ (HCG 50b). The
radial velocities of the
two subgroups differ by $\sim 900$ km s$^{-1}$, which if viewed as a
radial separation corresponds to a distance between two subgroups of
about 13 Mpc. 
Regarding the group HCG 82, we assume that it
consists of 6 accordant redshift galaxies, $b$, $c$, and the SDSS
galaxies 3, 4, 5 and 6 with $\langle cz \rangle =10550$ km s$^{-1}$ 
and $\sigma_v$=233 km s$^{-1}$. We suppose that two galaxies, $a$ 
and $d$, with mean $\langle cz \rangle =11430$ km
s$^{-1}$ and $\sigma_v \simeq 250$ km s$^{-1}$, are background galaxies 
either completely unrelated to the group or they are infalling to the
group. The radial velocity difference between the group and the
possible background 
double is again $\sim900$ km s$^{-1}$, with corresponding possible 
radial separation of 13 Mpc\footnote{Note that including the original groups
in our list change only slightly the results of the forthcoming
analysis (see Fig. 3).}.

Our final sample consists of 23 groups (since HCG 50 is split into 2
groups) which almost all have in their environment accordant
redshift galaxies. The groups HCG~35, HCG~58, HCG~60 and HCG~68
seem to be condensations within relatively rich groups 
consisting of 21, 15, 16 and 15 members respectively. 
Accordant redshift galaxies were not found in the 
environment of only one group, HCG~43.
Note also that accordant redshift galaxies were found by 
de Carvalho et al. (1997) and Zabludoff \& Mulchaey 
(1998) in the nearest environment of 15 more HCGs. Thus, most of HGGs, 
together with the accordant redshift galaxies in their near
environment, contain between 4 and $\sim 20$ galaxies, while
their projected length reaches $\sim$ 1-2 Mpc.

\section{HCG and Loose Group association}

\subsection{Alignment of the HCG with its environment}
We find that the projected distribution
of the accordant redshift galaxies lie in most cases along the
position angle defined by the original HCG galaxies. 
In Figure 2 we present  
the distribution of relative position angles, $\delta \phi$,
between the original HCG position angle and that of the environment
galaxies (in which case the original HCG is weighted only by its 
centre of mass). 
Similar alignment effects have been found between neighboring clusters
(e.g. Binggeli 1982, Plionis 1994) as well as between the major axis of
the brightest clusters galaxies and that of the parent cluster itself
(e.g. Plionis et al. 2003 and references therein).
It is evident that accordant redshift
galaxies are distributed preferentially along the elongation of the
HCG (see also Tovmassian \& Tiersch 2001; Tovmassian 2001), 
implying group formation by infall and accretion along large-scale filaments.

The significance of the observed alignment signal is estimated by applying 
a Monte-Carlo procedure by which we ask what is the
probability that at least half of the 16 $\delta\phi$ values, randomly selected
between 0 and 90 degrees, will have $\delta\phi<22\fdg5$, as
it observed in Figure 2. We find that this probability is only 0.025.
\begin{figure}[htb]
\includegraphics[width=90mm]{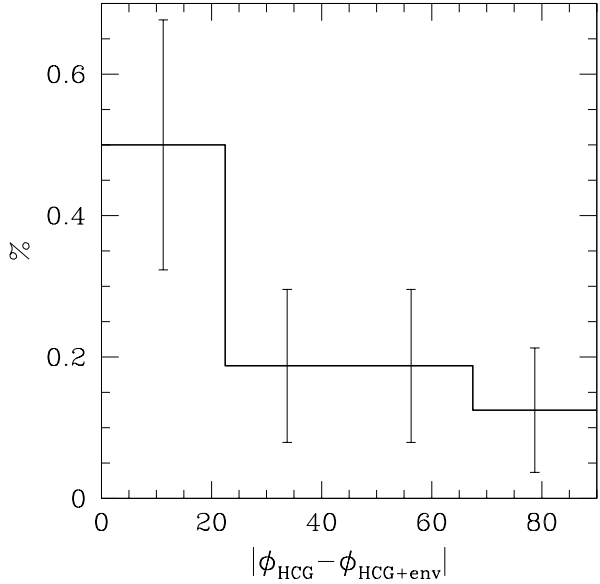}
\caption{The distribution of misalignment angle between the HCG and
its environment, where the latter includes the centroid of the HCG.}
\end{figure}
This probability is also corroborated by estimating the first moment
of the fourier transfrom of the $\delta\phi$ distribution and
comparing it with the expected random moment distribution. This approach
gives a random probability of $\sim 0.029$, in rough agreement with our
Monte-Carlo procedure.

\subsection{Velocity dispersion and the shape of HCGs and their environment}
In Table 1 we present the different parameters of the original HCGs and the
corresponding values of the HCG+environment systems. 
We determined the axial ratio $q$ using the moment of inertia method 
(cf. Basilakos et al. 2000 and references therein; Plionis, 
Basilakos \& Tovmassian 2004). 
The HCG+environment systems with $N_m\le 10$ have $\langle q \rangle \simeq
0.29\pm0.09$, similar to the mean elongation found by Plionis Basilakos, \&
Tovmassian (2004) for the corresponding
UZC-SSRS2 group catalogue (USGC; Ramella et al 2002) which
is based on the Updated Zwicky Catalogue (UZC; Falco et al. 1999) and
the Southern Sky Redshift Survey (SSRS2; da Costa et al 1998). 
Hence, we may assume that similar to poor groups they have a
prolate-like shape (see also Malykh \& Orlov 1986).

Inspecting Table 1 we see that the velocity dispersion of HCG+environment 
systems does not differ significantly from
that of the original HCGs. 
In some HCG+environment systems their $\sigma_v$ is even smaller than that of the 
original CGs, supporting the suggestion (Tovmassian \& Chavushyan
2000, Tovmassian 2002) that LG members are gravitationally bound to 
corresponding CGs. Thus, HCGs appear to be condensations in richer and
gravitationally bound galaxy aggregates.

We also compare the mean axial ratio for all the
HCG+environment systems (which is $\langle q \rangle =0.44^{+
  0.10}_{-0.23}$) with that of the USGC groups. Due to the possible
intrinsic correlation of the group elongation with group membership,
we weight the original USGC $q$ values such that they respect the
membership distribution of the HCG+environment systems. Doing so we
find a USGC group axial ratio of $\langle q \rangle =0.436 \pm 0.09$
which is in complete agreement with that of the HCG+environment systems. 

{\scriptsize

\begin{table}[htb]
\caption[]{Parameters of CGs and HCG+environment systems:
In Col. 1 the HCG designation is given, in Col. 2 - the number
of members of the original HCG group, in Col. 3 the velocity 
dispersion of the original group,
in Col. 4 - the number of members of the HCG+environment 
system, in Col. 5 the velocity dispersion of the latter 
system and in Col. 6 the projected axial ratio of the fitted
spheroid.}
\label{tab1}
\tabcolsep 8pt
\begin{tabular}{lcccccc} \\

\hline
 HCG & $N_m$ & $\sigma_v$/km s$^{-1}$ & $N_m$ & $\sigma_v$/km s$^{-1}$  & $q$ \\
     & \multicolumn{2}{c}{HCG} & \multicolumn{3}{c}{HCG+environment} \\
\hline
 
7 & 4 &  89 &  7 &  89 & 0.22  \\
25 & 4 &  61 &  8 &  97 & 0.18  \\
35 & 6 & 350 & 21 & 306 & 0.63  \\
40 & 5 & 149 &  6 & 296 & 0.55  \\
43 & 5 & 222 &  5 & 222 & 0.61  \\
45 & 4 & 234 & 10 & 381 & 0.66  \\
49 & 4 &  34 &  9 &  60 & 0.19  \\
50 & 5 & 472 &  8 & 546 & -     \\
50a&   &  -  &  4 & 266 & 0.45  \\
50b&   &  -  &  4 &  96 & 0.16  \\
56 & 5 & 170 &  6 & 126 & 0.29  \\
58 & 5 & 162 & 15 & 140 & 0.46  \\
60 & 4 & 427 & 16 & 472 & 0.60  \\
63 & 3 & 132 &  7 & 188 & 0.21  \\
64 & 3 & 214 &  6 & 239 & 0.26   \\
66 & 4 & 306 &  9 & 414 & 0.78  \\
67 & 4 & 211 & 14 & 339 & 0.73  \\
68 & 5 & 172 & 15 & 155 & 0.65  \\
82 & 4 & 620 &  6 & 233 & 0.22  \\
87 & 3 & 121 &  6 & 215 & 0.19  \\
88 & 4 &  27 &  6 &  76 & 0.39  \\
92 & 4 & 390 & 10 & 337 & 0.53  \\
97 & 5 & 370 & 14 & 402 & 0.66  \\
98 & 3 & 120 &  8 & 238 & 0.30  \\
\hline
\end{tabular}
\end{table}
}

\subsection{Correlation of velocity dispersion, number of group
  members and shape}
Compact groups as well as poor groups of galaxies have been shown to
have prolate-like shapes 
(Malykh \& Orlov 1986; Orlov et al. 2001; Oleak et al. 1995; Plionis
et al. 2004), while their observed $\sigma_v$ appears to depend on the 
orientation  of the group with respect to the observer
(e.g. Tovmassian, Martinez, \& Tiersch 1999). 
The similarity of the axial ratio distribution and its mean value
between the original HCGs and the sample of 
HCG+environment  systems studied here and in the work of Tovmassian \&
Chavushyan (2000) suggests that the HCG+environment systems also have a
prolate-like shape.

\begin{figure}[t]
\includegraphics[width=99mm]{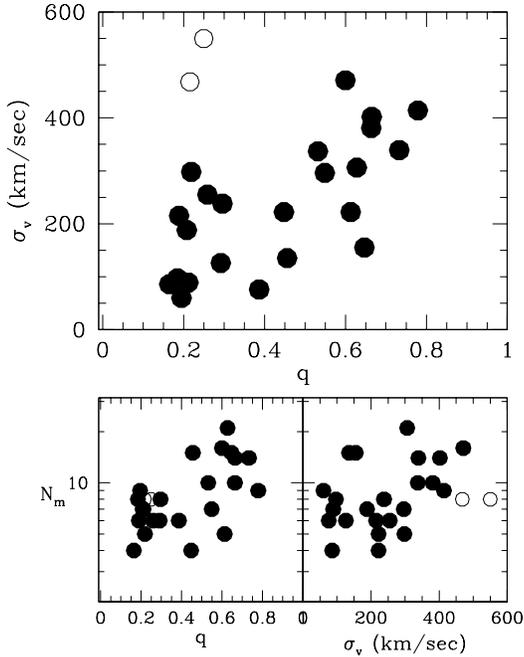}
\caption{Correlation between $\sigma_v$ and $q$ (upper panel), $q$ and
  $N_m$ (lower left panel) and $\sigma_v$ and $N_m$ (lower right 
panels). The
  open symbols represent the HCG+environment  groups containing the original
  HCG50 and 82, assumed to be the result of projection effects (see
  section 3.1). However, the $q-\sigma_v$ correlation persists even if we
  include these two groups in their original form, but it is weaker
  ($R=0.44$ and $P_{\rm random}=0.03$).}
\end{figure}

Therefore, if galaxies move in radial orbits
along the major axis of these highly elongated groups, one should
expect on the basis of projection effects
a correlation between the velocity dispersion of the groups and
their axial ratio, by which groups oriented close to 
the orthogonal to the line of sight will have relatively 
small projected $q$ and small
$\sigma_v$ while groups oriented close to the line of sight will have
larger $q$'s and larger $\sigma_v$'s.
  
Indeed, for the considered HCG+environment systems we find a strong 
($R=0.72$) and significant ($P_{\rm random}\simeq 10^{-4}$) 
correlation between $\sigma_v$ and $q$ (left panel of Fig. 3) with:
$$\sigma_v/{\rm km \; s^{-1}} = 403 (\pm 91) q +  61 (\pm 44) \;.$$
This effect can also be seen if we divide the groups into those 
with $\sigma_v\ge 220$ km/sec (13 groups) and those with
$\sigma_v<220$ km/sec (10 groups) which have median $q$-value 
and 67\% and 33\% quantile limits of
$0.6^{+0.01}_{-0.15}$ and $0.21^{+0.01}_{-0.02}$, respectively.
 
However, beyond the orientation effect discussed, 
another effect could contribute 
or even dominate the observed correlation.
This is related to the possible different evolutionary stage of
groups with different values of $q$. Richer systems (with larger
$N_m$) will be relatively more virialized, having a higher value of $\sigma_v$,
and will therefore have a higher value of $q$, 
since virialization tends to sphericalize dynamical systems.
The central panel of Fig. 3 
shows that groups with larger $N_m$ indeed have a 
higher axial ratio $q$, with quite a strong correlation
($R=0.57$, $P_{\rm random}=0.005$).
Since $\sigma_v$ is also only weakly ($R=0.43$, $P_{\rm
  random}=0.05$) correlated with $N_m$
(lower right panel of Fig. 3), different stages of
virilization cannot be the sole reason for the
$\sigma_v-q$ correlation and thus the orientation effect should be 
a major contributer.

As a further test of the relative strength of the two effects
(orientation and virialization) we study the corresponding correlation
of the 17 groups with $N_m\le 10$, which are relatively poor
and thus are not expected to be at a very different dynamical stage.
Indeed, as can be seen in Fig.4 (lower left and right panels) 
for these groups we find no $q-N_m$ or $\sigma_v-N_m$
correlations ($R \simeq 0.25$ and $P_{\rm random}\simeq 0.3$ for
both), while we do find a strong $q-\sigma_v$ correlation
(upper panel of Fig.4) with $R=0.69$ and $P_{\rm random}=0.002$. 
\begin{figure}[t]
\includegraphics[width=99mm]{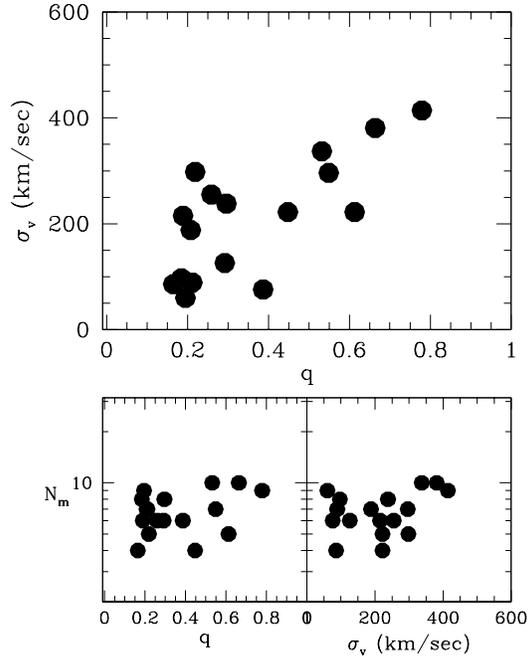}
\caption{As in Fig.3 but for the poorest of the HCG+environment groups
($4\le N_m \le 10$).}
\end{figure}

However, we would like to caution the reader that it is also possible
that groups of even equal mass could be
at different dynamical stage, depending on the environment in
which they are located. Indeed, Ragone \& Plionis (2006, in preparation),
using large cosmological N-body simulations, have found that
there is a significant correlation between the velocity dispersion 
and the group elongation (in 3D space but which survives also in 
projection) for equal mass group halos.

Finally, we investigate whether discreteness effects can introduce a
bias in the correlations discussed so far.
Tracing ellipsoids with a discrete distribution of particles can bias 
the intrinsic shape of the parent structure (e.g. Paz et al. 2006). 
Projection from three to two dimensions also
alters the intrinsic axial ratio distribution (cf. Hubble 1926; Carter
\& Metcalfe 1982) with the projected structures appearing more
spherical and typically smaller than the three dimensional ones.
We have investigated the coupling of the two effects by Monte-Carlo
simulations. We trace a 3 dimensional prolate spheroid of
predetermined axial ratio and size with a variable number of
particles. We then project the distribution of particles in the three
Cartesian planes after randomly orienting the spheroids with respect
to a predetermined line of sight. 
\begin{figure}[t]
\includegraphics[width=90mm]{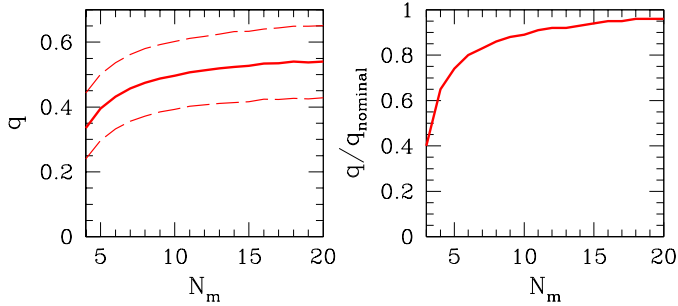}
\caption{Left panel: The projected mean axial ratio of Monte-Carlo 
prolate groups with an intrinsic mean axial ratio of $0.45$ as a function 
of points used to sample the group. Right panel: The corresponding
fractional decrease of the projected axial ratio
as a function of the number of points used to sample the group}.

\label{fig5}
\end{figure}

In the left panel of Figure 5 we present the median projected $q$ values 
together with the 67\% and 33\% quantile limits, 
derived from our Monte-Carlo analysis, as a function of the number of 
points used to sample the 3D prolate spheroid. We have used 
an intrinsic axial ratio for the 3D prolate spheroid of
$\beta=0.43$ which provides a projected axial ratio of $\sim 0.5$,
 near the value of our best sampled (richer) HCG+environment groups.

It is evident that there is a 
trend of decreasing sphericity with decreasing group galaxy
membership as in the observed group case.
In the right panel of Fig.5 we show the
fractional decrease of the nominal projected group axial ratio, which in turn
provide us with a multiplicative correction factor by which we can attempt to
crudely correct the observed $q$ group values for the effect of
discreteness. Note that small variations in the input nominal group
$q$-value changes insignificantly the fractional correction factor.

We quantify again the previous correlations but using now the corrected
group $q$ values and we find that the $q-\sigma_v$ correlation remains
mostly unaltered and significant ($R=0.63$ and $P_{\rm
  random}=0.001$), while the $q-N_m$
correlation, which could be viewed as that resulting from the different
virialization stages of groups of different richness, becomes weak and
completely insignificant ($R=0.25$ and $P_{\rm  random}=0.26$).

These results indicate that the orientation effect could be the dominant one in
shaping the $q-\sigma_v$ correlation. As a final point we note that
the dynamical properties and correlations
of HCG+environment systems do not differ from those of poor groups. 

\subsection{The $M_K$ magnitudes of E,S0 galaxies in HCGs}
Tovmassian, Plionis, \&  Andernach (2005 hereafter TPA) found that the
mean K-band absolute magnitude, $\langle M_K \rangle$, of E/S0
galaxies in USGC poor groups, located at distances $1000<cz<5500$ km 
s$^{-1}$, is equal to $-23\fmag42$. This is on average brighter by $0\fmag75$ than 
that of isolated E/S0 galaxies and of spiral galaxies (Sa and later) either in 
groups or in the field (in the same redshift range and
within the same - within 0.2 mags - apparent magnitude limit). 
Furthermore, the $\langle M_B \rangle$ values 
of isolated E/S0 and spiral galaxies practically do not differ 
from each other. This fact allowed TPA to conclude that E/S0's in
groups are formed as
the result of the merging of two galaxies of comparable luminosity and
thus mass (if we assume the same $M/L$).

In order to perform a similar study using Compact Groups, we have selected all
HCGs that lie in the previously mentioned redshift range, and we determined 
absolute K-band magnitudes, as in TPA, by using 
the $K_{s-total}$ magnitudes from $2MASS$ (see Jarrett et
al. 2000; {\it http://www.ipac.caltech.edu/2mass}), corrected for
Galactic extinction according to Schlegel, Finkbeiner, \& Davis
(1998). Furthermore, the group radial velocities
were corrected for the peculiar velocity of the 
Local Group and a local velocity field that contains a Virgo-centric infall 
component and a bulk flow given by the expectations of linear theory (see 
Branchini, Plionis, \& Sciama 1996).

Note that we do not restrict our HCGs sample to that
of Table 1 because most of these HCGs have redshifts outside the range
of interest. We extend our sample to include all HCGs that fall
within this range (in total 20 groups) and therefore in the estimation
of the mean absolute galaxy luminosities we use only HCG galaxies and not accordant
redshift galaxies of the environment
(which we have identified only for the HCGs which
fall within the SDSS survey limits).

It is important to note that the magnitude limits of the galaxy
samples from which the USGC and HCG groups have been selected, are
different ($m_B\sim 15.5$ and $\sim 17$, respectively). 
This implies that the two galaxy samples will trace a different fraction of
intrinsically faint and bright galaxies. The HCG galaxies being drawn 
from a fainter sample than the USGC sample will 
contain a larger fraction of nearby faint galaxies. 
However, we can overcome the bias introduced by: 
(a) excluding from our comparison the HCG
galaxies with apparent magnitudes larger than the USGC limit ($m_B\sim
15.5$) and (b) limiting our analysis to the same redshift range and 
requiring that the two samples have statistically 
compatible redshift distributions. 
\begin{figure}[t]
\includegraphics[width=80mm]{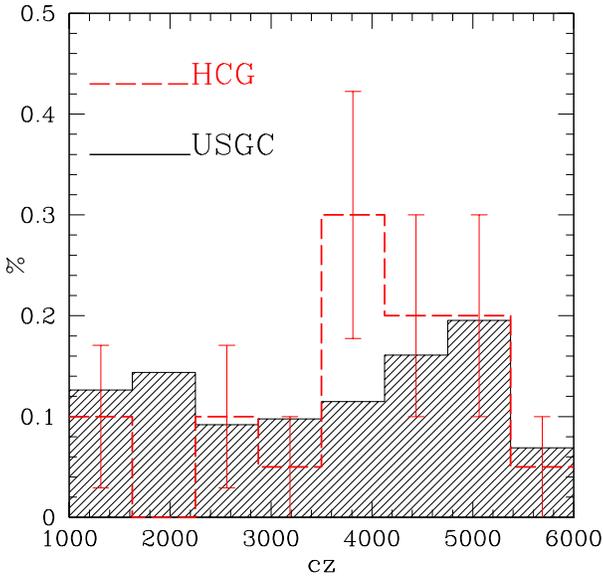}
\caption{The redshift distribution of the USGC and HCG groups in the
  range $1000<cz<5500$ km/sec.}
\end{figure}
In Figure 6 we present the two group
redshift distributions and indeed, we verify that this
is the case since the KS two-sampled test shows that the probability
of the two redshift distributions been drawn from the same parent
distribution is 0.15. 

In Table 2 we compare the $\langle M_K \rangle$ 
values for the USGC and HCG groups within the same radial
velocity limits and with $m_B\mincir 15.5$.
It is evident that the corresponding average absolute magnitudes
of early type galaxies between the USGC and HCG groups are almost identical 
($\delta M_K\sim 0.04$), from which
we infer that in both HCG and USGC groups, the E/S0 galaxies 
are formed similarly by the merging of two spiral galaxies of 
roughly the same luminosity (see TPA).
\begin{table}[htb]
\caption[]{The mean absolute magnitudes $\langle M_K \rangle$ of E/S0 and  
spiral galaxies in HCG and poor USGC groups. The quoted uncertainty is
  the distribution's standard deviation. The galaxy numbers are shown
in the parenthesis.}
\label{tab2}
\tabcolsep 10pt

\begin{tabular}{lccc} \hline
 & E/S0 & Spirals \\ \hline
USGC & $-23.83\pm1.18$ (156) &  $-22.96\pm1.25$ (310)  \\
HCG  & $-23.79\pm1.12$ (24)  & $-22.72\pm1.39$ (24) \\ \hline
\end{tabular}
\end{table}

We note that a more complete study of the possible
luminosity differences of the two group samples would entail the
derivation of their respective luminosity functions (eg. Mendes de
Oliveira \& Hickson 1991; Kelm \& Focardi 2004), which is however out of
the scope of the present study.

\subsection{The fraction of E/S0 galaxies}
TPA also found that the mean fraction of E/S0 galaxies
of poor USGC groups with $4\le N_m\le 10$ is 
$\langle f_{E/S0}\rangle \sim 0.23$, significantly higher than that of 
the Karachentseva, Lebedev \& Shcherbanovskij (1986) sample of
isolated galaxies, which is equal to $\sim 0.15$. They also showed 
that $f_{E/S0}$ very weakly depends on the group richness,
while a significant, although 
weak, correlation exists between the group $f_{E/S0}$ and velocity
dispersion.

In this work we use the morphological classification
of HCG member galaxies from Hickson (1994). Note that we have no such
classification for the accordant redshift galaxies. Therefore we
will only consider the HCG galaxies in this section.
We caution the reader that this may not 
properly represent the morphological content of the 
whole HCG+environment system. 

The mean fractions of E/S0 galaxies in the HCG groups is $\langle 
f_{E/S0}\rangle \sim 0.43$ (with the corresponding median of 0.4) 
which is about twice the value of the USGC poor groups, which in 
turn is $\sim$ 65\% higher than that of isolated galaxies (Tovmassian, 
Plionis \& Andernach 2004). Kelm \& Focardi (2004) have also found 
that the UZC compact groups have a higher 
content of E/S0's than that found among isolated galaxies. 
They also showed that the 
fraction of E/S0's among close neighbours of compact groups is 
intermediate between that of CGs and isolated galaxies.

Furthermore, we find a weak correlation between HCG+environment
richness ($N_m$) and $f_{E/S0}$ (with $R=0.36$ and random probability
$P=0.11$; see left panel of Fig.8).
For example, the median value $f_{E/S0}$ 
of groups with $N_m\leq 8$ is equal to $0.33^{+0.07}_{-0.03}$ while
the corresponding value of the groups with $N_m\geq 9$ is 
$0.40^{+0.10}_{-0.15}$.
This correlation, albeit weak, seems to be in agreement with
the well-established morphology-density relation (e.g. Dressler 1980;
Goto et al. 2003; Helsdon \& Ponman 2003). 

We also find a weak correlation between $\sigma_v$ and 
$f_{E/S0}$ with $R=0.37$ and $P_{\rm random}=0.1$ (see right panel of Fig.8). 
For example, dividing the HCG+environment groups in two subsamples
based on their velocity dispersion we find,
for those with $\sigma_v \geq 220$ km s$^{-1}$ a value of 
$\langle f_{E/S0}\rangle=0.40^{+0.10}_{-0.15}$
and for those with $\sigma_v \leq 220$ km s$^{-1}$,
$\langle f_{E/S0}\rangle=0.25^{+0.08}_{-0.15}$. 

Using the sample of richer LGs in the environment of some poor 
groups and HCGs, Zabludoff \& Mulchaey (1998) also found that 
$f_{E/S0}$ is definitely correlated with $\sigma_v$. 
\begin{figure}[t]

\includegraphics[width=85mm]{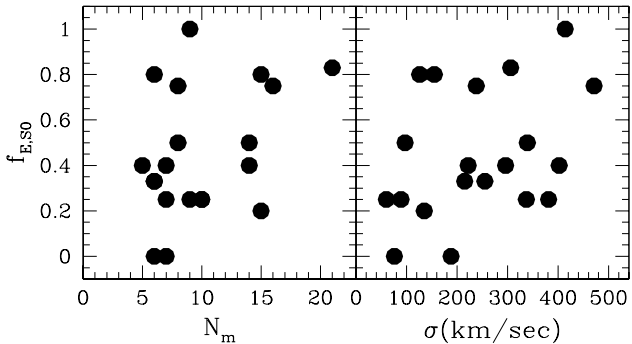}
\caption{Correlation of the fraction $f_{E/S0}$ with $N_m$ (left
  panel) and with $\sigma_v$ (right panel) for the HCG+environment
  groups.}
\end{figure}
In order to explain the $f_{E/S0} - \sigma_v$ correlation they
suggested that either galaxy morphology is set by the size of 
the local potential at the time of galaxy formation (Hickson, 
Kindl \& Huchra 1988) and/or that $\sigma_v$ and $f_{E/S0}$ 
increase as a group evolves (Diaferio, Geller, \& Ramella
1995).
The correlation, albeit weak, between $f_{E/S0}$ and the HCG X-ray luminosity (Ponman et
al. 1996) also suggest that groups with high values of
$f_{E/S0}$ are dynamically more evolved, possibly virialized systems.
 
In line with this model, Tovmassian, Plionis, \& Andernach (2004) suggested
that in elongated prolate-like groups, in which members move
predominantly along the group elongation, the probability of 
encounter and merging, and thus formation of E/S0 galaxies, is quite
higher than in a spherical overdensity of the same major axis size in which
galaxy orbits are ``chaotic''.
In effect what we suggest is that in such highly elongated groups, and prior to
virialization, galaxies move along the group elongation in
semi-regular orbits, as expected during the anisotropic infall of
galaxies in groups/clusters along filaments. During this phase and due
to the initially lower velocity dispersion of these unrelaxed systems,
many interactions and merging takes place (from which the higher
fraction of E/S0 emerges). Subsequently, the velocity dispersion of
these systems increase due to virialization processes which also
results in heating the intergalactic gas, which then emits in X-rays.

\subsection{The fraction of active galaxies}
The high density environment of the HCGs in combination with their relative
low velocity dispersion constitutes these environments as probable
sites of strong galaxy interactions, suggested also by their higher
fraction of E/S0 galaxies. If the AGN fueling mechanism is
triggered by such interactions one would expect to see a different
fraction of AGNs with respect to the field. 
Kelm et al. (1998) find that $\sim
3\%$ of HCG members are Seyferts, while if one includes LINERs this
fraction increases to $\sim$18\% (Coziol et al 2000). Furthermore, the
latter authors showed that including low-luminosity systems this
fraction increases even further to $\sim 40$\%, in agreement with
Ho et al. (1997).

For our sample of 14 HCGs we have spectral classification for 32 HCG
galaxies and 87 close environment galaxies. Since we do not have the
spectral classification of all the galaxies in the HCGs and their
environment, we stack together all the available data in two samples
(HCG and close environment galaxies). Furthermore, the 
different distances of the HCGs analysed in this work,
produce a variable sampling of the individual luminosity function of the
different HCG and their environment. Therefore, the following analysis 
is mostly valid as a comparison between the activity in HCGs and 
their close environment.

It is interesting that no Seyfert 1 spectra was identified, 
in agreement with the analysis of
Coziol et al. (2000) of the Southern CG sample (Iovino 2002). However, in our case,
we cannot exclude that small number statistics could be the cause of such a paucity.
Most importantly, we find 
31\% ($\pm 10\%$) and 13\% ($\pm 6\%$) of AGNs and starburst galaxies, 
respectively, in the HCGs, while these fractions are 29\% ($\pm 6\%$) and 4.6\% 
($\pm 2.3\%$) for the close environment galaxies. If we restrict
our analysis to Seyfert type AGNs (excluding LINERs) we find that
their fractions are 6.5\%
and 8\% respectively between the HCGs and their close environment. 
These fractions are in general agreement with Coziol et al (2000) but the
Seyfert fraction appears to be higher than that of Kelm et al. (2004).  

The most important results of this analysis is that
\begin{itemize}
\item the fraction of AGNs (being Seyferts or LINERs)
does not vary significantly between the HCGs and the looser groups within which they
are embedded (see also Shimada et al. 2000), and
\item what varies at a significant level appear to be
the fraction of starbursting galaxies, which is quite higher in the
dense environment of the HCGs. 
\end{itemize}
The former could be the result of
perturbations caused on disk galaxies by the HCG tidal field, even
before close galaxy encounters and interactions take place in the HCG centres. 
Such perturbations can cause an inflow of gas towards the disk centre,
triggering bursts of star formation and possibly feeding the AGN 
(e.g. Byrd \& Valtonen 1990, Fujita 1998).
This could explain why
the fraction of AGNs is the same in the HCGs and in their close
environment, implying that the AGNs are well in place already in the
looser groups. 
The higher fraction of starbursting galaxies in the HCGs with respect
to their close environment,
then implies that only starburst activity is 
enhanced by the close encounters and
interactions that take place in the dense HCGs.

\begin{figure}[t]
\includegraphics[width=80mm]{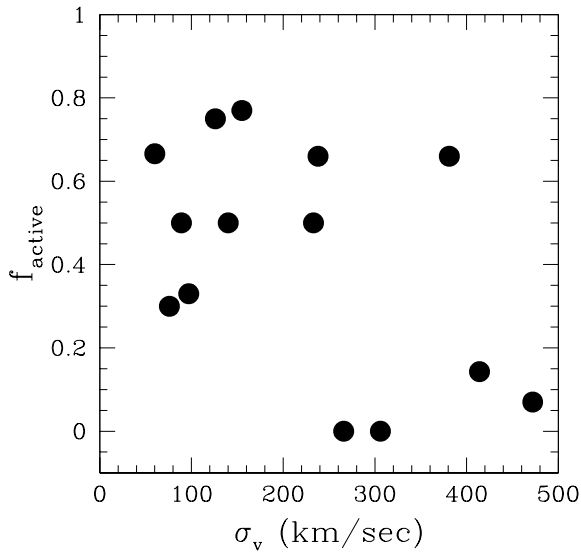}
\caption{Correlation of the fraction of active HCG+environment
  galaxies, $f_{active}$, with the HCG+environment velocity
  dispersion.}
\end{figure}
Furthermore, Focardi \& Kelm (2002) and 
Coziol et al. (2004) have found a correlation between the dynamics of
the CG and the galaxy member activity, with galaxies being more active
and dominated by late-type morphologies 
in lower velocity dispersion systems. We verify their result also in our
the HCG+environment systems, as can be seen in Figure 9, were we correlate the fraction of
active galaxies (AGNs and starbursts) with the HCG+environment
velocity dispersion. An anti-correlation is present with Spearman
correlation coefficient of $R=-0.45$ and the probability of random
correlation being $P=0.12$. This anti-correlation is robust to
the exclusion of individual groups.

So a possible overall picture emerges in which high velocity
dispersion compact groups, which are also X-ray luminous
(e.g. Ponman et al 1996), are dynamically more evolved and 
their late-type galaxy members had
enough time to interact, merge and evolve into earlier types. The
low-velocity dispersion CGs, are dynamically younger, richer in
late-type and in active galaxies, yet unevolved.

\section{Conclusions}
Using SDSS spectroscopic data we have searched the close
environment of 22 HCGs and found that the
large majority is embedded in galaxy overdensities that extend up to 
$\sim$1 Mpc around the HCG centre.
A wealth of indications point in the direction that
HCGs are condensations within looser prolate-like groups of galaxies with 
dynamical and morphological characteristics similar to those
of poor groups (like the USGC groups
studied by Plionis et al. 2004).

For example, (a) the HCG+environment elongation
is similar to that of the USGC
groups, (b) the HCG+environment groups show the same axial ratio - velocity
dispersion correlation as the USGC groups, a correlation which, as we have
shown, could be due to an orientation effect of intrinsically 
prolate-like groups,
(c) the mean K-band absolute magnitude $\langle M_K \rangle$ of E/S0 
galaxies in HCGs is about the same as that in poor USGC groups, which is 
brighter by $\sim 0\fmag75$ than that of isolated E/S0's and group
or isolated spiral galaxies, showing that E/S0's in CGs, as in USGC 
groups, could be formed as the result of the merging of two galaxies of
similar luminosity,
(d) the mean fraction of early type galaxies 
in HCGs is higher than in the field but also twice as high as in poor 
USGC groups, indicating enhanced merging in the central compact
condensations of poor groups, 
(e) the fraction $f_{E/S0}$ depends weakly on the
richness and the velocity dispersion of the HCG+environment groups
(f) the fraction of AGNs is similar between members of
 HCGs and their close environment galaxies, while the
fraction of starburst galaxies is significantly higher in the HCGs,
(g) the fraction of active galaxies (AGNs
and starbursts) is anti-correlated with the velocity dispersion of the
HCG+environment systems.

These results could be explained by noting that groups are very 
elongated prolate-like structures (e.g. Plionis, Basilakos \& Tovmassian
2004) in which galaxies should move in radial orbits around the 
group gravitational centre. 
Thus we expect that in many occasions what appears to be a 
compact group is just an ordinary group, or part of it, observed
either when some of its member galaxies happen to be close to each other, 
during their radial orbits along the group elongation, or if the 
group is oriented close to the line of sight, and all its members are 
projected over a small solid angle. Of course in the former case
gravity will play its role in inducing interactions, the results of
which will depend on a multitude of initial conditions (relative
galaxy velocities, orbit trajectories, local background density,
galaxy morphology and gas content, etc). Our results indicate that the
main outcome of such interactions, in the compact group centre,
is an enhanced starburst activity but not necessarily 
the triggering of an AGN.

Furthermore, our results show that the members of 
groups with high velocity dispersion are predominantly 
early-type non-active galaxies while members of presently low-velocity groups
(probably indicating their earlier evolutionary stage) are predominantly
late-type active galaxies. These findings are in good agreement with
Coziol, Brinks \& Bravo-Alfaro (2004) and suggest
an evolutionary sequence by which initially late-type galaxies, in an
initially low velocity dispersion group (indicating either a low-mass
system or a dynamically unevolved more massive system), interact with the
group tidal field - by which both the AGN and 
starburst activity is triggered - then
virialization processes, which increase the group velocity dispersion and
X-ray emission, and galaxy merging will result 
into a high velocity dispersion, X-ray luminous, group with a 
higher fraction of early-type galaxies. Note also, that numerical
simulations (G\'{o}mez-Flechoso \& Dominguez-Tenreiro 2001)
have shown that there is a lack of enhanced merger activity in the
quiescent virialized CG halos suggesting that early-type galaxies form
in the initial group formation stage, when the CG velocity dispersion
is still low.

\begin{acknowledgements}
HMT and MP acknowledge funding by the Mexican Government grants
SEP-2003-C02-44376/A-1 and CONACyT-2002-C01-39679, respectively. We
thank the referee, Gary Mamon, for many useful comments and
suggestions.
\end{acknowledgements}

\end{document}